\documentclass[fleqn,usenatbib]{mnras}

\usepackage[T1]{fontenc}
\usepackage{ae,aecompl}

\usepackage{amsmath,color}
\usepackage{graphicx}
\usepackage{amssymb}

\newcommand{\new}[1]{#1}
\newcommand{\hide}[1]{}
\renewcommand{\d}{\delta}

\title{On the non-Poissonian repetition pattern of FRB121102}

\author[N. Oppermann et al.]{Niels Oppermann,$^{1,2}$\thanks{E-mail: niels@cita.utoronto.ca}
Hao-Ran Yu,$^{1,3}$
Ue-Li Pen$^{1,2,3,4,5}$
\\
$^{1}$Canadian Institute for Theoretical Astrophysics, University of Toronto, 60 St.\ George Street, Toronto ON, M5S 3H8, Canada\\
$^{2}$Dunlap Institute for Astronomy and Astrophysics, University of Toronto, 50 St.\ George Street, Toronto ON, M5S 3H4, Canada\\
$^{3}$Tsung-Dao Lee Institute, Shanghai Jiao Tong University, Shanghai, 200240, China\\
$^{4}$Canadian Institute for Advanced Research, 180 Dundas St.\ West, Toronto ON, M5G 1Z8, Canada\\
$^{5}$Perimeter Institute for Theoretical Physics, 31 Caroline St.\ North, Waterloo ON, N2L 2Y5, Canada
}

\date{Accepted XXX. Received YYY; in orignal form ZZZ}

\pubyear{2016}

\begin{document}
\label{firstpage}
\pagerange{\pageref{firstpage}--\pageref{lastpage}}
\maketitle

\begin{abstract}
	The Fast Radio Burst FRB121102 has been observed to repeat in an irregular fashion. Using published timing data of the observed bursts, we show that Poissonian statistics are not a good description of this random process. As an alternative we suggest to describe the intervals between bursts with a Weibull distribution with a shape parameter smaller than one, which allows for the clustered nature of the bursts. We quantify the amount of clustering using the parameters of the Weibull distribution and discuss the consequences that it has for the detection probabilities of future observations and for the optimization of observing strategies. \new{Allowing for this generalization, we find a mean repetition rate of $r=5.7^{+3.0}_{-2.0}$ per day and index $k=0.34^{+0.06}_{-0.05}$ for a correlation function $\xi(t)=(t/t_0)^{k-1}$}.
\end{abstract}


\section{Introduction}
\label{sec:introduction}

Fast Radio Bursts (FRBs) are bright, short-duration (ms), pulses detected in the radio waveband. They are characterized by high dispersion measures ($10^2$--$10^3\,\mathrm{pc}\,\mathrm{cm}^{-3}$) and their origin is not yet understood. Several bursts have been observed to arrive from a single location and with the same dispersion measure, collectively referred to in the following as FRB121102. None of the other FRBs have been observed to emit repeated pulses. However, it is possible that these have merely escaped detection.

The bursts of FRB121102 seem to happen in an irregular fashion, but appear to be clustered to some degree. This clustering may contain hints to the emission mechanism. Additionally, it has important consequences for the strategy that will have to be followed in order to detect the maximum number of bursts in the future and for the maximum repeat rate that one may infer from non-detections of repetitions for other FRBs, as discussed by \citet{connor-2016}.

In this paper, we suggest to model the distribution of intervals between successive bursts of FRB121102. This allows us to consider simple deviations from Poissonian statistics and to quantify the clustering.  Any clustering parameter estimation explicitily or implicitly requires a Bayesian prior to quantify error bars on the inferred parameters.
To this end, we discuss the Weibull distribution in the next section. We introduce the data we use in Sect.~\ref{sec:data}, derive the statistical formalism in Sect.~\ref{sec:formalism}, and discuss our results and their consequences in Sect.~\ref{sec:results}. \new{A brief summary of our findings and their consequences is given in Sect.~\ref{sec:summary}.}

\section{Beyond Poissonian statistics}
\label{sec:weibull}

For a Poissonian point process, a sufficient statistic is the total number of events. Here we consider deviations from purely Poissonian statistics and make use of the additional information given by the distribution of events.

If the bursts of FRB121102 were distributed in time according to a Poisson point process with a constant expectation value, then the distribution of intervals $\d$ between subsequent bursts would be exponential, i.e.,
\begin{equation}
	\mathcal{P}(\d|r) = r\,\mathrm{e}^{-\d\,r},
\end{equation}
where $r$ is the constant rate of bursts. One possible generalization of this distribution of intervals is the Weibull distribution,
\begin{equation}
	\label{eq:weibull}
	\mathcal{W}(\d|k,r) = k\d^{-1} \, \left[\d \, r\, \Gamma\left(1 + 1/k\right)\right]^k \, \mathrm{e}^{-\left[\d \, r \, \Gamma\left(1 + 1/k\right)\right]^k},
\end{equation}
which has an additional shape parameter $k$. Here,
\begin{equation}
	\Gamma(x) = \int_0^\infty \mathrm{d}t \, t^{x - 1} \, \mathrm{e}^{-t}
\end{equation}
is the gamma function. We choose a parameterization in which the parameter $r$ can again be interpreted as a rate, since
\begin{equation}
	r^{-1} = \left<\d\right>_{(\d|k,r)} = \int_0^\infty \mathrm{d}\d \, \d \, \mathcal{W}(\d|k,r).
\end{equation}
For $k=1$ the Weibull distribution reduces to the Poissonian case. For values of $k$ different from 1, the Weibull distribution describes a certain degree of clustering. Specifically, for $k < 1$, small intervals are favored compared to the Poissonian case and thus the presence of a burst makes an additional burst in the near future more likely.

With this parameterization, we can ask whether the existing observations of bursts and observations with an absence of bursts are consistent with purely Poissonian statistics and, if not, we can quantify how strong the clustering of bursts is.

\section{Data}
\label{sec:data}

\begin{figure}
	\includegraphics[width=\columnwidth]{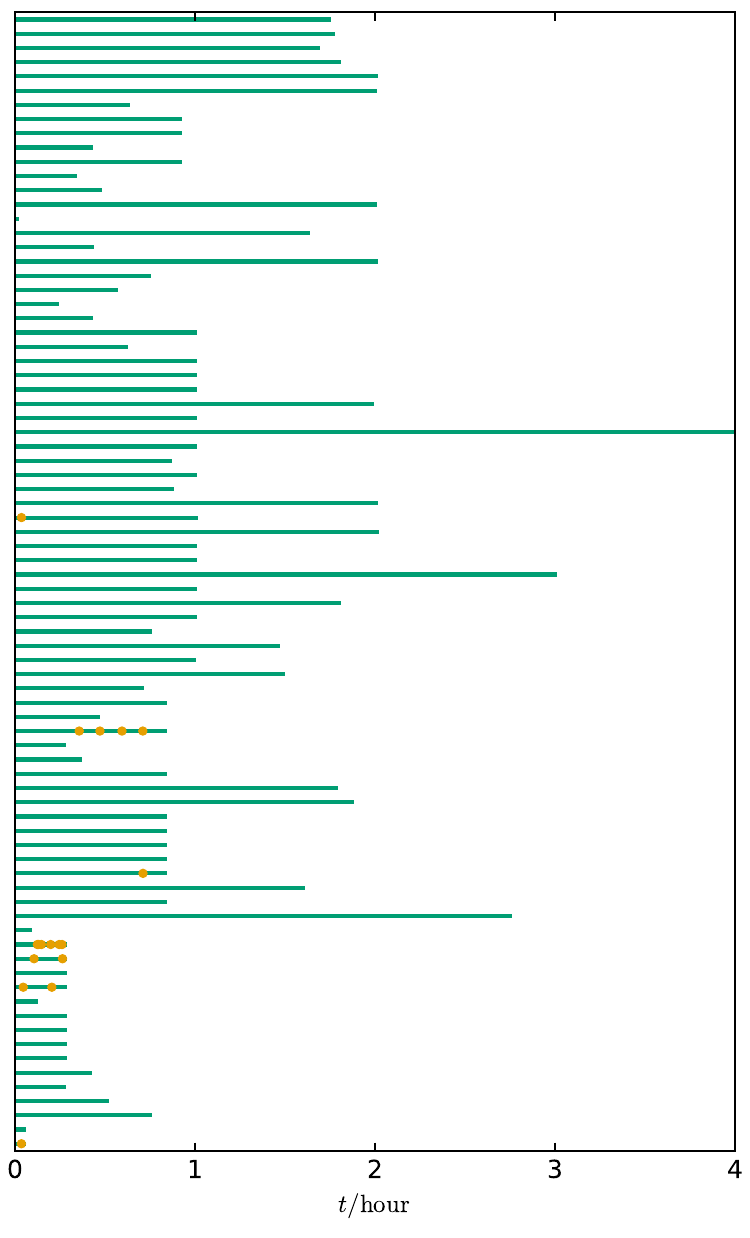}
	\caption{\label{fig:intervals}Lengths of observational intervals (\emph{green lines}) with positions of bursts within them (\emph{orange dots}). \new{The vertical ordering of the intervals is chronological with the most recent observation at the top.} All intervals have been shifted to the same start time.}
\end{figure}

For this analysis, we use the bursts listed in Table~1 of \citet{spitler-2016} and Table~3 of \citet{scholz-2016}, as well as the starting and stopping times of the observations listed in Table~2 of \citet{scholz-2016}. To get all times onto equal footing, we convert the starting and stopping times to barycentric modified Julian dates and apply a frequency-dependent dispersion correction. These corrections have already been applied to the published burst arrival times and we apply them here to the observation intervals to obtain the correct time differences. The lengths of the resulting observational intervals and the location of each burst within these intervals are shown in Fig.~\ref{fig:intervals}.

\new{Further follow-up observations of FRB121102 have been conducted and more bursts have been detected \citep[e.g.,][]{chatterjee-2017}. However, the exact duration of each observation is not published and thus we do not include these in our analysis. While more data will constrain the parameters more tightly, the qualitative conclusions of this study will not change once the necessary data from newer observations are included.}

\section{Formalism}
\label{sec:formalism}

\begin{figure}
	\includegraphics[width=\columnwidth]{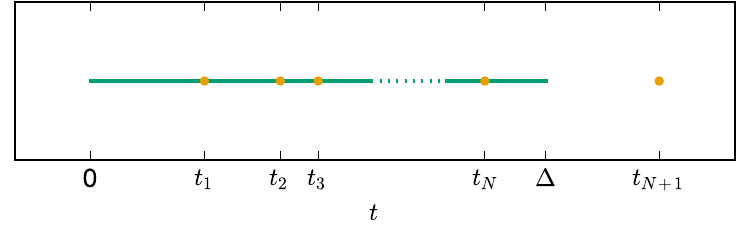}
	\caption{\label{fig:singleinterval}Illustration of the variables used to describe a single finite-duration observation (\emph{green line}). We assume that $N$ bursts (\emph{orange dots}) happen during the observation.}
\end{figure}

Consider first a single observation of a finite duration $\Delta$. The information from this observation consists of the number of observed bursts, $N$, and the times at which these bursts happened, $t_1, t_2,\dots, t_N$. For simplicity, we measure times with respect to the start time of the observation. Figure~\ref{fig:singleinterval} illustrates the variables we use.

We calculate the likelihood for $N$ and $t_1,\dots,t_N$ by marginalizing over the time of the next burst, $t_{N+1}$,\footnote{We use $\mathcal{P}$ to denote probability densities and $P$ to denote probabilities.}
\begin{align}
	\label{eq:likelihood_firstsplit}
	&\mathcal{P}(N,t_1,\dots,t_N|k,r)\nonumber\\
	&= \int_{t_N}^\infty \mathrm{d}t_{N+1} \, P(N|t_1,\dots,t_N,t_{N+1}) \, \mathcal{P}(t_1,\dots,t_N,t_{N+1}|k,r).
\end{align}
The first ingredient is the probability that $N$ bursts are observed, given the times of the $N+1$ first bursts. This is independent of the parameters of the Weibull distribution and trivially given by
\begin{align}
	P(N|t_1,\dots,t_N,t_{N+1}) &= \theta(\Delta - t_N) \, \theta(t_{N+1} - \Delta),
\end{align}
where $\theta(\cdot)$ is the Heaviside step function, which is 1 for a positive argument and 0 otherwise.

The second ingredient, the probability for \new{the arrival times of the $N+1$ first bursts}, given the parameters of the Weibull distribution, can be split into a product of probabilities for the inter-burst intervals,
\begin{align}
	\label{eq:likelihood_times}
	\mathcal{P}(t_1,\dots,t_N,t_{N+1}|k,r) & = \mathcal{P}(t_1|k,r) \, \prod_{i = 1}^{N - 1} \mathcal{P}(t_{i+1}|t_{i},k,r)\nonumber\\
	&= \mathcal{P}(t_1|k,r) \, \prod_{i = 1}^{N - 1} \mathcal{W}(t_{i+1} - t_i|,k,r).
\end{align}

The first factor in the last equation, the probability density for the time of the first observed burst, is slightly more complicated than the others. Since there is no previously observed burst, this is not a pure Weibull distribution. However, we will relate the two distributions in the following way: Consider the interval $\d_{01} = t_1 - t_0$ between the last unobserved burst and the first observed burst. The probability distribution for the length of this interval is characterized by the Weibull distribution, which describes the distribution for interval lengths in general, and the fact that the observation started within this interval. The probability for the latter fact is proportional to the length of the interval so that we can write
\begin{align}
	P(\textnormal{obs.\ start during }\d_{01} | \d_{01}) &\propto \d_{01}
\end{align}
and thus
\begin{align}
	&\mathcal{P}(\d_{01}|k,r,\textnormal{obs.\ start during }\d_{01})\nonumber\\
	&~~~~~\propto P(\textnormal{obs.\ start during }\d_{01}|\d_{01}) \, \mathcal{P}(\d_{01}|k,r)\nonumber\\
	&~~~~~\propto \d_{01} \, \mathcal{W}(\d_{01}|k,r).
\end{align}
To ensure the correct normalization, the full probability distribution has to be
\begin{align}
	\mathcal{P}(\d_{01}|k,r,\textnormal{obs.\ start during }\d_{01}) = \frac{\d_{01}}{\bar{\d}} \, \mathcal{W}(\d_{01}|k,r),
\end{align}
where
\begin{equation}
	\label{eq:dbar_weibull}
	\bar{\d} = \int_0^\infty \mathrm{d}\d \, \d \, \mathcal{W}(\d|k,r) = \frac{1}{r}.
\end{equation}

Further, for symmetry reasons, we have
\begin{align}
 \mathcal{P}(t_1|\d_{01}) = \frac{1}{\d_{01}} \, \theta(\d_{01} - t_1).
\end{align}
Thus, we can write the first factor in Eq.~\eqref{eq:likelihood_times} as
\begin{align}
	\mathcal{P}(t_1|k,r) &= \int_0^\infty \mathrm{d}\d_{01} \, \mathcal{P}(t_1|\d_{01}) \, \mathcal{P}(\d_{01}|k,r,\textnormal{obs.\ start during }\d_{01})\nonumber\\
	&= \frac{1}{\bar{\d}} \, \int_{t_1}^\infty \mathrm{d}\d_{01} \, \mathcal{W}(\d_{01}|k,r)\nonumber\\
	&= \frac{1}{\bar{\d}} \, \mathrm{CCDF}(t_1|k,r),
\end{align}
where we have defined the complementary cumulative distribution function
\begin{align}
	\label{eq:cdf_weibull}
	\mathrm{CCDF}(\d|k,r) &= \mathrm{e}^{-\left[\d\,r\,\Gamma(1 + 1/k)\right]^k}
\end{align}
in the last line.

Plugging everything into Eq.~\eqref{eq:likelihood_firstsplit}, we are left with a product of $N - 1$ Weibull distributions, a cumulative Weibull distribution due to the time difference $\Delta - t_N$ at the end of the observation, and another cumulative Weibull distribution due to the time elapsed at the beginning of the observation before the first burst. As argued above, this last factor has an additional factor $1/\bar{\d}$. In total, we obtain
\begin{align}
	\label{eq:likelihood_weibull}
	\mathcal{P}(N,t_1,\dots,t_N|k,r) =& ~\frac{1}{\bar{\d}} \, \mathrm{CCDF}(t_1|k,r) \, \mathrm{CCDF}(\Delta - t_N|k,r)\nonumber\\
	&\prod_{i=1}^{N-1} \mathcal{W}(t_{i+1} - t_i|k,r).
\end{align}
This expression is valid for any distribution of intervals. To obtain the specific expression for the case of the Weibull distribution, we plug in Eqs.~\eqref{eq:weibull}, \eqref{eq:dbar_weibull}, and \eqref{eq:cdf_weibull}.

Strictly speaking, Eq.~\eqref{eq:likelihood_weibull} is only valid for $N > 1$. For $N = 1$ it simplifies trivially to
\begin{align}
	\label{eq:likelihood_N1}
	\mathcal{P}(N=1,t_1|k,r) =& ~\frac{1}{\bar{\d}} \, \mathrm{CCDF}(t_1|k,r) \, \mathrm{CCDF}(\Delta - t_1|k,r).
\end{align}

For the case $N=0$, i.e., an observation without detected bursts, we follow a similar argument as above and marginalize over the time of the first burst after the start of the observations,
\begin{align}
	P(N=0|k,r) &= \int_\Delta^\infty \mathrm{d}t_1 \, \mathcal{P}(t_1|k,r)\nonumber\\
	&= \frac{1}{\bar{\d}} \, \int_\Delta^\infty \mathrm{d}t_1 \, \mathrm{CCDF}(t_1|k,r).
\end{align}
Plugging in Eq.~\eqref{eq:cdf_weibull} yields the result for the Weibull distribution,
\begin{align}
	\label{eq:likelihood_N0}
	P(N=0|k,r) &= \frac{\Gamma_\mathrm{i}{\left(1/k, \left(\Delta\,r\,\Gamma(1+1/k)\right)^k\right)}}{k\,\Gamma{\left(1+1/k\right)}},
\end{align}
where
\begin{align}
	\Gamma_\mathrm{i}(x,z) = \int_z^\infty \mathrm{d}t \, t^{x-1} \, \mathrm{e}^{-t}
\end{align}
is the incomplete gamma function.

This completes the derivation of the likelihood for a single finite-duration observation. In the limit that successive observations are widely spaced when compared to the spacing of the bursts, the individual observations become independent and the likelihood turns into a simple product of the likelihoods of the individual observations. In the following, we follow this approximation. In Appendix~\ref{sec:finiteness} we show that the error introduced by this simplification is small.

To infer the parameters $k$ and $r$ we need to multiply the likelihood with a prior and calculate the posterior distribution for the two parameters. Since both parameters are strictly positive we work with their logarithms. Specifically, we choose independent Jeffreys priors for $k$ and $r$, i.e.,
\begin{equation}
	\mathcal{P}(k,r) \propto k^{-1} \, r^{-1}
\end{equation}
\new{so that
\begin{equation}
	\mathcal{P}(\log k,\log r) = \mathrm{const}.
\end{equation}}
We then calculate the posterior as
\begin{align}
	\mathcal{P}(k,r|N,t_1,\dots,t_N) &\propto \mathcal{P}(N,t_1,\dots,t_N|k,r)\,\mathcal{P}(k,r).
\end{align}

\section{Results and discussion}
\label{sec:results}

\begin{figure}
	\includegraphics[width=1.0\columnwidth]{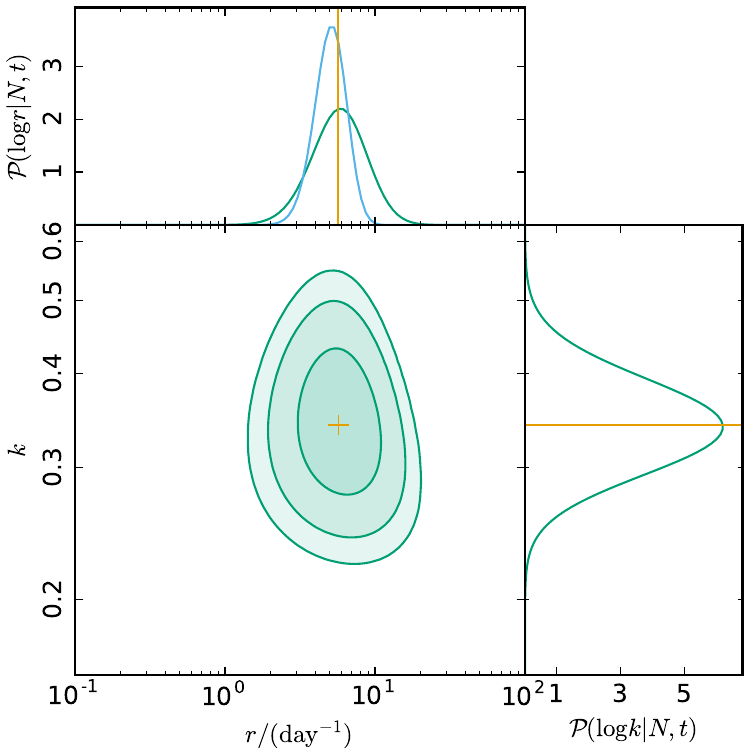}
	\caption{\label{fig:result}Posterior probability distribution for the two parameters of the Weibull distribution (\emph{green}). The contours correspond to the 68\,\%-, 95\,\%-, and 99\,\%-confidence regions. The dashed \emph{blue} curve in the top panel depicts the slice through the posterior at $k=1$, i.e., the posterior for the rate $r$ when restricting the Weibull distribution to a Poissonian distribution. The \emph{orange} lines indicate the posterior mean parameters.}
\end{figure}

We make use of the data shown in Fig.~\ref{fig:intervals} and insert them into Eqs.~\eqref{eq:likelihood_weibull}, \eqref{eq:likelihood_N1}, and \eqref{eq:likelihood_N0}. Figure~\ref{fig:result} shows the resulting two-dimensional posterior distribution for $k$ and $r$, as well as the one-dimensional marginal posteriors for the individual parameters.

\subsection{Non-Poissonian nature of FRB121102}
\label{sec:non-poissonian}

From Fig.~\ref{fig:result} we immediately see that the Poissonian case of $k=1$ is strongly disfavored by the data. The posterior mean value for the shape parameter is \new{$\left<k\right>_{(k|N,t)} = 0.34^{+0.06}_{-0.05}$} and the posterior mean for the rate is \new{$\left<r\right>_{(r|N,t)} = 5.7^{+3.0}_{-2.0}/\mathrm{day}$ The uncertainties given here are the 68\,\% confidence intervals. The estimated rate is similar to the Poissonian rate, but the uncertainty interval is about twice as wide.} Since the value of $k$ is less than 1, there is indeed an excess of short inter-burst intervals. A consequence of this is that, if a burst is observed, the probability that another burst happens shortly after is increased.

Although we have not exhausted all possible distributions to model the data, it can be quantitatively shown that the Weibull distribution describes the data much better than Poissonian case. We examine this by Kolmogorov-Smirnov (KS) test on the distribution of burst intervals. For Poissonian case, the distribution of intervals disagrees with the exponential best fit at level $\alpha=1.03\times 10^{-6}$. If we assume a Weibull distribution, we have to consider 3 kinds of intervals ($N=2$, $N=1$ and $N=0$ as discussed in \S \ref{sec:formalism}) separately. For each of these, the distribution depends on the Weibull parameters $k,r$, and additionally the distribution of observation durations (see Appendix~\ref{app:interval} for more details). These three distributions disagree with the Weibull model at levels $\alpha=1.46, 1.18\times 10^{-4}, 0.37$ respectively. It shows that Weibull model performs much better than Poissonian. Note that, although much better than Poissonian in $\alpha$ the Weibull distribution still does not describe the data well, especially in the distribution of ``$N=1$''-kind intervals. The reason is obvious -- from Fig.~\ref{fig:intervals} we see many clustered bursts to accumulate in short observations, and even very likely to be in the beginning or ending of one observation.

\begin{figure}
	\includegraphics[width=1.0\columnwidth]{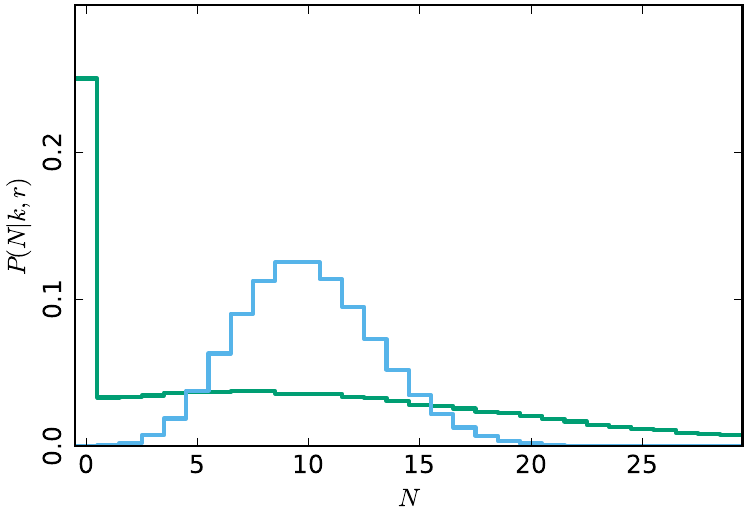}
	\caption{\label{fig:n_singleint}Probability to see $N$ bursts during an observation for which the expected number of bursts is 10. The solid \emph{green} histogram shows the Weibull case with parameters fixed to the posterior-mean values and the dashed \emph{blue} histogram shows the Poissonian case with the same rate.}
\end{figure}

To illustrate the effect this has on the number of bursts during an observation, we consider the case of a Weibull distribution with parameters fixed to the posterior-mean values. We numerically calculate the probability to see $N$ bursts during an observation that lasts ten times the mean burst separation, i.e., $\Delta = 10/r$. \new{For the posterior-mean repetition rate of FRB121102, this is a 42-hour observation. The result} is shown in Fig.~\ref{fig:n_singleint}. For comparison, we also plot the Poissonian probability. While the mean number of bursts is the same for both cases, the Weibull case has a much higher probability of yielding no bursts at all. The excess of short intervals makes it likely that either no burst is seen (if the observation happens to fall within a long inter-burst interval) or several bursts are seen, since the presence of one burst makes the presence of other bursts in the temporal vicinity more likely.

\begin{figure}
	\includegraphics[width=1.0\columnwidth]{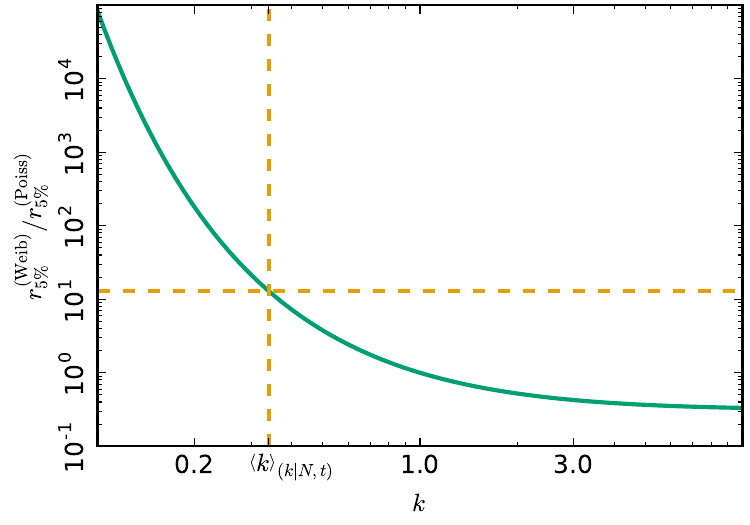}
	\caption{\label{fig:ruleout}\new{Ratio of the thresholds above which rates can be ruled out at the 95\,\% level for the Weibull case and the Poissonian case after conducting a continuous observation without detection. For $k = 1$ the ratio is 1 by definition. The vertical dashed line marks the posterior-mean value of $\left<k\right>_{(k|N,t)}=0.34$, for which we find a ratio of 13, indicated by the horizontal dashed line.}}
\end{figure}

\new{An important consequence is that the inference that can be done on the mean repetition rate from observing an FRB, or a random location in the sky, for some time $\Delta$ and not seeing any bursts is much weaker than in the Poissonian case. A possible line of reasoning is to equate the probability of seeing no bursts in such an observation with a probability threshold $\alpha$, to solve for the threshold rate $r_\alpha$, and to conclude that rates above this threshold value are disfavored by the observational evidence. For the Poissonian case, this leads to a threshold rate of
\begin{equation}
	r^{\mathrm{(Poiss)}}_\alpha = - \frac{\ln(\alpha)}{\Delta},
\end{equation}
whereas the Weibull result can be obtained by numerically solving Eq.~\eqref{eq:likelihood_N0} for $r$. Figure~\ref{fig:ruleout} shows the ratio of the threshold rate in the Weibull case, $r^{\mathrm{(Weib)}}$, and the threshold rate in the Poissonian case, as a function of the shape parameter $k$. For the posterior-mean value, we find a ratio of 13 for $\alpha = 5\,\%$, meaning that rates that are consistent with the data of such an observation can be 13 times as high when allowing for $k \neq 1$ than when assuming Poissonian statistics. Choosing a lower probability threshold $\alpha$ makes this ratio even larger.}

\new{Thus, it is possible within the Weibull model that an FRB that has been observed not to repeat for some time does indeed repeat with a fairly high average rate, but with a value of $k$ that is significantly less than 1. This opens the door to the possibility that all observed FRBs are in fact the same class of objects, even though only one object has been observed to burst repeatedly.}

\subsection{Clustering}
\label{sec:clustering}

\new{To further quantify the degree of clustering, we consider the power spectrum of the burst density for the Weibull distribution with the parameters fixed to the posterior-mean values. If we discretize the time axis into bins of width $\mathrm{d}t$, the mean number of bursts during a bin is
\begin{equation}
	\bar{N} = r\,\mathrm{d}t.
\end{equation}
We can calculate a density of bursts by dividing by the bin width, $n(t) = N(t)/\mathrm{d}t$, and use its Fourier-space version $n(\nu)$ to define the power spectrum,
\begin{equation}
	P(\nu) = \left< \left|n(\nu)\right|^2 \right>_{(N|k,r)}.
\end{equation}}

\begin{figure}
	\includegraphics[width=1.0\columnwidth]{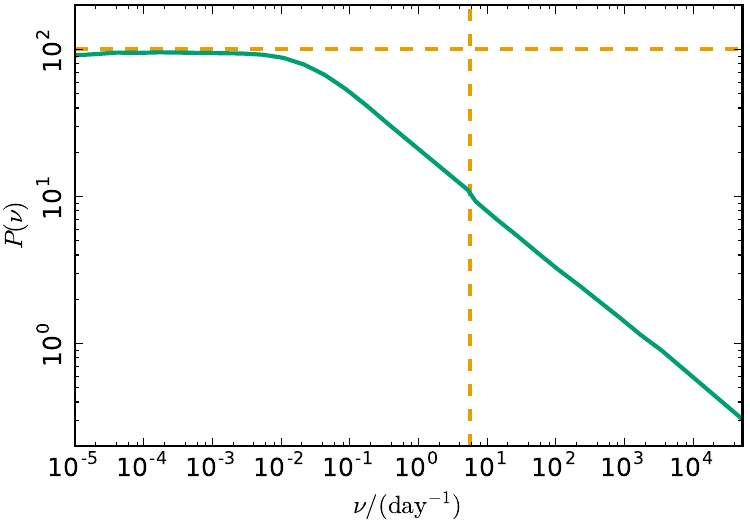}
	\caption{\label{fig:powspec}\new{Power spectrum of the burst number density for the Weibull distribution with parameters fixed to the posterior-mean values (\emph{green line}). The Poissonian expectation ($r=5.7$) has been subtracted to show the excess power due to clustering. The horizontal \emph{orange dashed} line shows the Poissonian power spectrum increased by a factor $(\Gamma(1 + 2/k)/\Gamma(1 + 1/k)^2 - 1)\sim 17.6$. The vertical \emph{orange dashed} line marks the mean rate $\left<r\right>_{(r|N,t)}$.}}
\end{figure}

\new{For a Poissonian point process, the power spectrum is flat, i.e., independent of $\nu$. We choose the normalization of the Fourier transform in such a way that the value of the power spectrum, except for the zero frequency, is the mean number of bursts per day. In Fig.~\ref{fig:powspec} we show the numerical result for the power spectrum of Weibull-distributed events after subtracting the Poissonian power, i.e., we show the excess of power due to the clustering. Two extremes are of interest: For frequencies much larger than the mean rate $r$, the clustering becomes unimportant and the Poissonian power spectrum is recovered. The clustering power declines as a power law in frequency. For frequencies much smaller than the mean rate the power spectrum becomes flat as well, but with an increase in power by a factor $(\Gamma(1 + 2/k)/\Gamma(1 + 1/k)^2 - 1)$. We derive this result in Appendix~\ref{app:power}. At intermediate frequencies, the power spectrum smoothly transitions between these two regimes and it is this range that describes the non-trivial clustering.}

\new{
A common parametrization of temporal clustering is through the Fourier
Transform of the power spectrum, called the two-point correlation function, $\xi(t) \equiv (2\pi)^{-1}\int_0^\infty
P(\nu) \, \exp(i t\nu)\, \mathrm{d}\nu$, which describes the fractional 
excess over Poissonian at time lag $t$. The asymptotic forms of the
correlation function at short and long lags are
}
\begin{equation}
\xi(t)=\left\{ \begin{array}{ll}  k \, \left[\, \Gamma\left(1 + 1/k\right)\right]^k  (rt)^{k-1} & t \lesssim 10/r \\
               \frac{ \Gamma(1 + 2/k)/\Gamma(1 + 1/k)^2 - 2}{rt} & t \gg 10/r\end{array} \right..
\end{equation}


%

\subsection{Observational strategies}
\label{sec:obs_strategies}

The fact that the bursts appear clustered should inform future observational campaigns that aim to detect more bursts. In the Poissonian case, the probability of detecting a burst depends only on the total duration of the observation. In the presence of clustering, however, an observational strategy that spreads the same observing time over a larger period by introducing gaps has a higher chance of detecting a burst.

\begin{figure}
	\includegraphics[width=1.0\columnwidth]{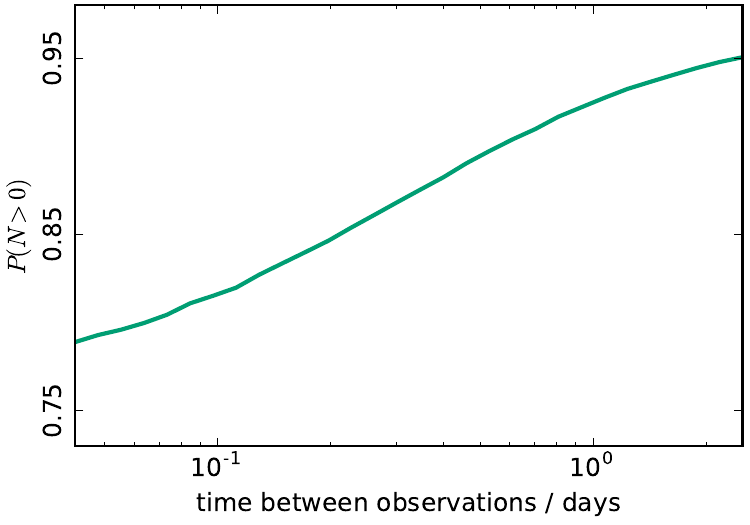}
	\caption{\label{fig:obs_strategy}Probability of detecting at least one burst as a function of observing strategy. The assumed campaign consists of ten \new{equally long} observations of \new{a duration such that the Poissonian expectation value for the number of bursts is 10}, interrupted by nine equally long gaps. The graph shows the detection probability as a function of the gap duration.}
\end{figure}

To illustrate this effect, we consider \new{again the example of an observation whose total duration is such that the expected number of bursts is 10. However, we now split this time into ten equally long observations with equally long gaps between them.} Figure~\ref{fig:obs_strategy} shows the numerically calculated probability of detecting at least one burst in such an observation, as a function of the length of each observing gap. Evidently, spreading the observations out increases the odds of a successful detection. Here we have again fixed the parameters to the posterior-mean values \new{and the Poissonian probability of detecting at least one burst is 99.995\,\%.}

Calculations such as the one shown here can be done for arbitrary observing strategies and may therefore be useful for their optimization.

\section{Summary}
\label{sec:summary}

\new{We have used the Weibull distribution as a generalization from Poisson statistics. This is not the only possible extension. Examples for other extensions could be a gamma distribution of intervals or a Poissonian model with time-varying rate, which could for example be modeled as a log-normal random field. We use the Weibull distribution as a simple, mathematically tractable, model that allows us to demonstrate that the available information on burst arrival times and the observation timing strongly disfavors Poissonian statistics and favors instead a clustered distribution. The most important consequences are:
\begin{itemize}
	\item For a continuous observation, a non-detection of bursts is much more likely than in the Poissonian case.
	\item Consequently, upper limits on repetition rates derived from non-detections become much looser and rate estimates in general less certain.
	\item A spread-out observation interspersed with gaps is better suited for detecting further bursts than continuous observations with the same total observation time.
\end{itemize}}

\appendix
\section{Effect of the finite time between observations}
\label{sec:finiteness}

In our analysis we treat each of the observational intervals shown in Fig.~\ref{fig:intervals} as independent. In reality, however, they are separated by finite gaps and the correlations in the burst pattern extend, at least in principle, across these gaps and thus correlate the observations across observational intervals. Here we investigate the severity of this effect.

We run two series of simulations of Weibull-distributed events, with parameters fixed to the posterior-mean values. In the first series, we simulate events in the intervals shown in Fig.~\ref{fig:intervals} assuming that they are completely independent, i.e., infinitely separated. In the second series, we simulate the intervals at their actual separation. We then apply the analysis described in Sect.~\ref{sec:formalism} to each simulated data set. If the finite separation of the observations had a great effect, we would expect to see posterior distributions that are different for the two simulation series and are shifted away from the simulated parameters in the case of the realistic simulation. However, neither appears to be the case for the two times 240 simulations that we have run.

\hide{To illustrate the similarity between the posterior distri- butions, we plot the average of the 240 posteriors for both types of simulations in Fig.~\ref{fig:finiteness}. The two mean posteriors are very similar, at least down to the 95\,\%-confidence level. We also note that the center of mass of both average posteriors is very close to the values of the parameters that were used in the simulations. Thus we see that our analysis is unbiased, even though we are neglecting the finite separation of the observations.

\begin{figure}
	\includegraphics[width=1.0\columnwidth]{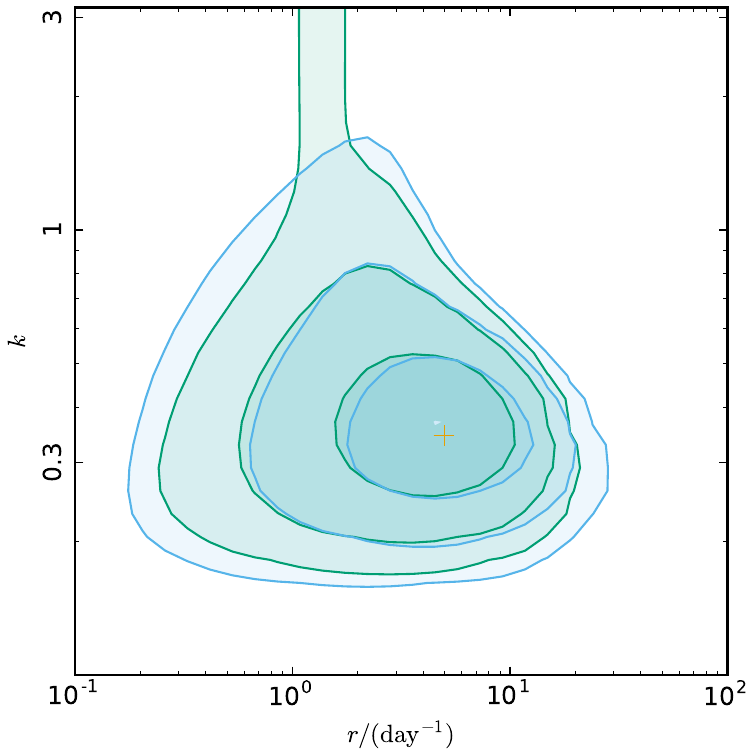}
	\caption{\label{fig:finiteness}Average posterior distributions over 240 simulations. The \emph{green} contours show the mean posterior in simulations that assume the observational intervals to be independent, whereas the \emph{blue} contours show the result of the same analysis applied to simulated data that takes the finite separation of the observations into account. The orange + marks the parameter values used in the simulations. \textbf{[TO DO: Still working on these sims; hopefully will come out a bit nicer with better statistics.]}}
\end{figure}
}
We check the accuracy of our model by computing the fraction of simulations for which the true parameter values lie within each contour. The result of this test is shown in Table~\ref{tab:contour_stats} and confirms that the size of our uncertainty region is accurate.

\new{As a null test, we have also run a similar set of simulations where we set $k=1$, i.e., the Poissonian case. In this case, we do find a posterior that is centered on $k=1$, thus confirming that the non-Poissonian structure we infer is not an artifact of our analysis.}

\begin{table}
\begin{center}
	\caption{Fraction of the simulations with independent intervals (\emph{second column}) and those with finite observational separations (\emph{third column}) for which the parameter values assumed in the simulations lie within the 68\,\%-, 95\,\%-, and 99\,\%-confidence regions derived from the simulated data.}
	\label{tab:contour_stats}
	\newdimen\digitwidth
	\setbox0=\hbox{\rm 0}
	\digitwidth=\wd0
	\catcode`*=\active
	\def*{\kern\digitwidth}
\begin{tabular}[tmb]{ccc}
	\hline
	confidence level& independent& finite\cr
	\hline
	68\,\%& 59\,\%& 63\,\%\cr
	95\,\%& 93\,\%& 93\,\%\cr
	99\,\%& 98\,\%& 98\,\%\cr
	\hline
\end{tabular}
\end{center}
\end{table}

\section{Probability density functions of intervals}

\label{app:interval}
We consider a number of events (e.g. FRB bursts here) detected by an independent series of observations with finite times. The duration of each observation $T_{\rm obs}$ obeys $f_{\rm obs}$. In the case of $T_{\rm obs}\rightarrow\infty$, we can observe the intrinsic statistics of intervals between two events -- we denote it $f^\infty_2(\Delta t)$. Realistically, when $T_{\rm obs}$ is finite, we observe 3 kinds of intervals -- combinations between observation boundaries and events. We define $t_0$ as time of interval between two boundaries, $t_1$ as time of interval between one event and one boundary, and $t_2$ as time of interval between two events. We study the distribution of $\{t_0,t_1,t_2\}$  given $f^\infty_2$ and $f_{\rm obs}$.

One can easily proof that the PDF of $t_0$ can be written as $f_0\propto P_0(t) f_{\rm obs}(t)$, where $P_0$ is the probability of no detections within $T_i$. For discrete cases, $f_{\rm obs}=N^{-1}\sum_{i=1}^{N} \delta(t-T_i)$, where $\delta(t)$ is the Dirac delta function. It is straightforward to integrate and get the CCDF of $t_0$,
\begin{equation}\label{eq.1-F0}
	1-F_0(t)=\frac{\sum_{i=m}^{N}P_0(T_i)}{\sum_{i=1}^{N}P_0(T_i)},
\end{equation}
where $m$ is the smallest index s.t. $T_m>t$.

To produce $t_1=t$, we need 
condition(1): Given an event, the probability that the next/previous event happens at least $t$ after/before it\footnote{condition(1)$\,\sim\,$condition(1$'$): Given a boundary, the next/previous event happens $t$ away from it.} $P_1(t)=1-F^\infty_2$ \footnote{In Eq.(\ref{eq.1-F0}), $P_0(t)=1-F^\infty_1(t)$.}, and condition(2): the probability that $T_i>t$, which is $1-F_{\rm obs}(t)$. Thus,
\begin{equation}
	f_1\propto f^\infty_1(t)\int_t^\infty f_{\rm obs}(t')dt' = f^\infty_1(t) (1-F_{\rm obs}(t)),
\end{equation}

To produce $t_2=t$, we need the probability of $\Delta t=t$ selected from $f^\infty_2(\Delta t)$, the probability of $T_i>t$, and the probability that $t$ can be put in $T_i$, which is $T_i-t$. So
\begin{equation}
	f_2\propto f^\infty_2(t)\int_t^\infty f_{\rm obs}(t')(t'-t)dt'=f^\infty_2(t)\sum_{i=m}^{N}(T_i-t),
\end{equation}
where again $m$ is the smallest index s.t. $T_m>t$.

Recall that for Weibull distribution,
\begin{align}
	& f^\infty_2=kt^{-1}[r\,\Gamma(1+1/k)\,t]^k \exp\left[-(r\,\Gamma(1+1/k)\,t)^k\right],\\
	& f^\infty_1 = P_1 = 1-F^\infty_2 = \exp\left[-(r\,\Gamma(1+1/k)\,t)^k\right],\\
	& P_0= \frac{\Gamma_\mathrm{i}{\left(1/k, \left(r\,\Gamma(1+1/k)\,t\right)^k\right)}}{k\,\Gamma{\left(1+1/k\right)}}.
\end{align}

\section{Power spectrum of a Weibull distribution}
\label{app:power}

\new{For a Weibull distribution of inter-burst intervals, the power spectrum on temporal scales much smaller than the mean burst separation and on scales much larger than this separation is flat. Here we derive the ratio of power between these two regimes.}

\new{Consider the time elapsed until $N$ successive events have occurred, $\Delta_N$. For large $N$ this will correspond to large time-scales and thus small frequencies. Since the inter-event intervals are considered to be independent, the central limit theorem holds and the distribution of $\Delta_N$ is Gaussian for large $N$. The mean and variance of this Gaussian are straightforwardly calculated from the mean and variance of an individual interval. The mean is
\begin{equation}
	\bar{\Delta}_N = \frac{N}{r}
\end{equation}
both for the Poissonian and the Weibull case. The variance is
\begin{equation}
	\sigma^2_{\Delta_N,\mathrm{Poiss}} = \frac{N}{r^2}
\end{equation}
for the Poissonian case and
\begin{equation}
	\sigma^2_{\Delta_N,\mathrm{Weib}} = \frac{N}{r^2} \, \left[\frac{\Gamma\left(1 + 2/k\right)}{\Gamma\left(1 + 1/k\right)^2} - 1\right]
\end{equation}
for the Weibull case.}

\new{Now consider a fixed time-scale $\Delta$. For large $\Delta$, the number of bursts that occur during this time, $N_\Delta$ will again be approximately Gaussian distributed with a mean
\begin{equation}
	\bar{N}_\Delta = \Delta \, r
\end{equation}
and a comparatively small variance, which is proportional to the variance of the duration of $\bar{N}_\Delta$ events, given above. We thus obtain as the ratio of the large-scale power in the Weibull case and the Poissonian power the factor
\begin{equation}
	\frac{P^{\mathrm{Weib}}(\nu \ll r)}{P^{\mathrm{Poiss}}(\nu \ll r)} = \frac{\Gamma\left(1 + 2/k\right)}{\Gamma\left(1 + 1/k\right)^2} - 1.
\end{equation}
At high frequencies, the Poisson noise subtracted power spectrum scales as $\propto \nu^{-k}$.
The numerical result shown in Fig.~\ref{fig:powspec} confirms the validity of this argument.}

\new{Note that the line of reasoning followed here is applicable more generally than just to the case of a Weibull distribution. For any distribution of independent intervals for which the central limit theorem has validity, i.e., any distribution with finite non-zero moments, a similar result can be derived.}

\section*{Acknowledgements}

We thank \new{Christina Peters and Liam Connor for helpful discussions and comments on the manuscript}. This research has made use of NASA's Astrophysics Data System. The figures were produced using the \texttt{matplotlib} library \citep{hunter-2007}. \new{Some of the results have been derived using the \texttt{NIFTY} package \citep{selig-2013}.} We acknowledge NSERC support.

\bibliographystyle{myaa}
\bibliography{weibull}

\bsp
\label{lastpage}
\end{document}